\documentclass[preprint2]{aastex}
\usepackage{amsmath,amssymb}
\usepackage{mathrsfs}
\usepackage{subfigure}

\slugcomment{Not to appear in Nonlearned J., 45.}

\shorttitle{CAN WE PROBE THE LORENTZ FACTOR OF GRBs?}
\shortauthors{Aoi et al.}

\begin{document}

\title{Can we probe the Lorentz factor of gamma-ray bursts from GeV-TeV spectra integrated over internal shocks?}

\author{Junichi Aoi\altaffilmark{1}, Kohta Murase\altaffilmark{2, 3}, Keitaro Takahashi\altaffilmark{4}, Kunihito Ioka\altaffilmark{5} and Shigehiro Nagataki\altaffilmark{1}}

\altaffiltext{1}{YITP, Kyoto University, Kyoto, 606-8502, Japan}
\altaffiltext{2}{Department of Physics, Tokyo Institute of Technology,
2-12-1, Ookayama, Meguro-ku, Tokyo, 152-8550, Japan}
\altaffiltext{3}{Center for Cosmology and AstroParticle Physics, 191 
West Woodruff Avenue, Columbus, OH 43210, USA;
murase.2@mps.ohio-state.edu}
\altaffiltext{4}{Department of Physics and Astrophysics, Nagoya University, Nagoya, 464-8602, Japan}
\altaffiltext{5}{KEK (High Energy Accelerator Research Organization), Tsukuba, 305-0801, Japan}

\begin{abstract}
We revisit the high-energy spectral cutoff originating from
the electron$-$positron pair creation in the prompt phase of gamma-ray bursts (GRBs)
with numerical and analytical calculations. We show that the conventional exponential 
and/or broken power law cutoff should be drastically modified to a
shallower broken power-law in practical observations that integrate emissions from 
different internal shocks.
Since the steepening is tiny for observations, this "smearing" effect can generally reduce the 
previous estimates of the Lorentz factor of the GRB outflows. 
We apply our formulation to GRB 080916C, recently detected by the Large Area Telescope 
detector on the \textit{Fermi} satellite, and find that 
the minimum Lorentz factor can be $\sim
600$ (or even smaller values), which is below but consistent with 
the previous result of $\sim 900$.
Observing the steepening energy (so-called "pair-break energy") is crucial
to diagnose the Lorentz factor and/or the emission site
in the future observations, especially current and future Cherenkov telescopes such as MAGIC, VERITAS, and CTA.
\end{abstract}


\section{Introduction}
\label{sec:introduction}
Gamma-ray burst (GRB) is one of the most mysterious objects in the universe. 
The typical energy of its prompt emission is $\sim 100-1000$ keV, which corresponds to the breaking point of the broken power law 
describing the energy spectrum well \citep{1993ApJ...413..281B}. Although there have been many works on the mechanism of 
prompt emission since its discovery, it is still unknown. 
Combined with the optically thin synchrotron emission mechanism, the internal shock model has been most widely discussed to 
explain the prompt emission \citep[see, e.g.,][for reviews]{2006RPPh...69.2259M,2007ChJAA...7....1Z,
2008FrPhC...3..306F}. 
In fact, this model can reproduce the behavior of complicated light curves 
\citep[e.g.,][]{1997ApJ...490...92K, 1998MNRAS.296..275D, 2007A&A...466...93M, 2009A&A...498..677B}.
The complicated light curve consists of several pulses and these pulses include sub-pulses which 
vary violently. 
Collisions among inhomogeneous outflows lead to the shock formation that converts the kinetic energy of the outflow 
to the internal energy. We can interpret the non-thermal gamma-ray emission as the emission of electrons 
which are accelerated at the shocks, where inhomogeneous outflows characterized by multiple sub-shells can easily 
produce highly variable light curves.

High-energy emission from GRBs was already detected by EGRET before \textit{Fermi}, but it has not been clear where the end point of the 
high-energy tail is \citep[e.g.,][]{1992A&A...255L..13S,1994ApJ...422L..63S,1994Natur.372..652H,1995ApJ...453...95S}.  
If the spectrum extends to higher-energy range, sufficiently high-energy photons cannot avoid the 
electron$-$positron pair-
creation process. Thus, it has been expected that there should be a cutoff due to this process, 
although there was no crucial observational evidence on it. 
The pair-creation cutoff energy generally depends on the emission radius $r$ and the 
bulk Lorentz factor of the outflow $\Gamma$. Hence, we can obtain information on these quantities once it is observed. 
Therefore, many authors investigated the possibility to extract the information from the cutoff 
energy \citep[e.g.,][]{1997ApJ...491..663B, 2001ApJ...555..540L, 2004ApJ...613.1072R, Murase:2007ya, 2008MNRAS.384L..11G, 2008ApJ...677...92G}. 

However, these studies focused on the high-energy emission from one emission zone, neglecting the time evolution of 
physical quantities during the emission. 
In realistic observations, it is likely that the observed emission comes from many emission regions, and 
physical quantities should be time dependent even during one sub-pulse produced by one emission zone.
Hence, the cutoff behavior in spectra should be affected by both (1) the time-evolution effect during one sub-pulse 
and (2) the time-integration effect over many sub-pulses. 
On the former point (1), the time variation of properties on the photon field during the sub-shell-crossing timescale 
($\delta t/(1+z) \sim l/c$) is significant and comparable to that during the total duration of the emission 
\citep{2008ApJ...677...92G, 2009A&A...498..677B}, which can alter the resulting high-energy spectrum. 
The cutoff also evolves reflecting the time evolution of the emission site and the Lorentz factor of a sub-shell \citep{2006ApJ...650.1004B}. 
The latter point (2) is also important, since the observed high-energy spectrum should be different from 
that expected in the one-zone treatment unless we can separate the contribution of each emission zone.   
More specifically, in the internal shock model, we would see the superposition of the emission from 
many collisions among different sub-shells during the total duration.
Especially, if a flux of a single sub-pulse is small, we cannot avoid integrating the contribution of all the 
sub-pulses to observe the burst. 
In such a case, it is important to study the effect of a lot of internal collisions. 

In this paper, we focus on the above time-integration effect over many sub-pulses in the internal shock model, and  
examine whether we can extract information about the bulk Lorentz factor of GRBs from high-energy GeV-TeV spectra. 
We model the high-energy emission from the multiple emission regions caused by multiple collisions, 
and calculate energy spectra numerically. We need to consider a lot of sub-shells with various Lorentz factors
to explain complicated and irregular light curves. Our numerical approach can treat this situation well that is 
valid even when the variance of the distribution of sub-shells' Lorentz factors is large. 
This is an advantage over the analytical approach which is valid only when the variance is small enough 
\citep[see, e.g.,][]{2010ApJ...709..525L}. In fact, the variance should be large in the internal shock model, 
in order that the sub-shells' kinetic energy is efficiently converted into the radiation energy as prompt emission 
\citep{1997ApJ...490...92K, 2001ApJ...551..934K}. 
For calculations, we simplify the emission from each collision using a rather phenomenological procedure. 
This may be too simple but enough to study the smearing effect by many internal collisions. 
Other observational effects coming from radiation mechanisms will be discussed later.  

In Section \ref{sec:radiation}, we explain the calculational method used in this work. We consider multiple sub-shells 
and integrate the energy spectrum of emission which comes from each merged shell. In Section \ref{sec:spectrum}, we show the 
results on the energy spectrum of GRBs. In Section \ref{sec:discussion}, we discuss the possibility to extract information 
about the Lorentz factor of the GRB outflow from observations. 
Moreover, we discuss the case of GRB080916C, which is an interesting event to study 
\citep{2009MNRAS.396.1163Z, 2009ApJ...698L..98W, 2009ApJ...700L..65Z}. 
We also compare our numerical results with the analytical results.
In Appendices A and B, we explain the dynamics of shells which approximate the inhomogeneous outflows of GRBs. 

\section{Method}
\label{sec:radiation}
The inhomogeneous outflow is characterized by multiple sub-shells. For simplicity, we only consider completely inelastic 
collisions between shells and neglect the interaction via pressure waves (which is a good approximation if the outflow energy is 
dominated by the kinetic part as shown in \citep{2000A&A...358.1157D}). The kinetic energy of the shells is converted to the 
radiation energy via electron acceleration in the internal shock model, where 
we may phenomenologically give the energy spectrum of emission 
from each merged shell (see Section \ref{subsec:radiation}). 
In Appendices A and B, we summarize the dynamics of multiple collisions which was similar to that given in 
\cite{1997ApJ...490...92K}. 
As described in Section \ref{subsec:pair}, we can evaluate the cutoff energy of the spectrum, which originates from 
the electron$-$positron pair-creation process. 
Hereafter, we express physical quantities that are defined in the comoving frame of the 
outflow with primes (e.g., photon energy $\varepsilon'$) and those which are defined in the source (or laboratory) frame 
without primes 
(e.g., $\varepsilon$). 

\subsection{Radiation from a Merged Shell}
\label{subsec:radiation}
The internal energy $E_{\rm{int}}$ is determined when two shells merge (see the Appendix A). This energy is divided into 
the energy of protons, electrons and magnetic fields. 
The parameter $\epsilon_{p}$ indicates the fraction of 
the energy in (thermal and nonthermal) protons. Similarly, the energies of (nonthermal) electrons and magnetic fields 
are characterized by $\epsilon_{e}$ and $\epsilon_{B}$, respectively.
For simplicity, we just assume $\epsilon_{e} = \epsilon_{B} = \epsilon_{p} = 1/3$. In addition, 
we consider the fast cooling case, where almost all the energy of electrons accelerated in each merged shell is released as 
radiation \citep[e.g.,][]{gg03}.

Observed energy spectra of prompt emission are usually fitted well by broken power-law ones 
\citep{1993ApJ...413..281B}. 
This broken power-law spectrum is characterized
by low-energy power-law index $\alpha$, 
high-energy power-law index $\beta$, and the break energy $\varepsilon_{b}$. 
Therefore, we 
assume that the energy spectrum for each collision is expressed by the 
broken power law,
as expected for the synchrotron and/or inverse-Compton emission 
from nonthermal electrons. 
Although there may be an additional component in the high-energy spectrum \citep[e.g.,][see also later discussions in Section \ref{subsec:other}]
{2003Natur.424..749G}, 
we approximate the spectrum using the broken power-law spectrum 
for the demonstrative purpose. 
We here adopt $\alpha =1, \beta = 2.2$, and $\varepsilon_{b} = 300$ keV, which are typical values obtained 
from observations \citep[e.g.,][]{2000ApJS..126...19P}. Given the spectral shape,
we can normalize a photon distribution function $dn/d\varepsilon$ by
the total radiation energy $\epsilon_{e} E_{\rm{int}}$ with
\begin{equation}
\epsilon_{e} E_{\rm{int}} = (1+z) V 
\int d \varepsilon \varepsilon \frac{dn}{d\varepsilon},
\label{eq:eint}
\end{equation}
where $V$ is the volume of each emission region (see the Appendix A). 
The distribution function of photons in the source frame is given by
\begin{equation}
\frac{dn}{d\varepsilon} = \begin{cases}
							K \left( \frac{\varepsilon}{\varepsilon_{b}} \right)^{-\alpha}  & \text{$ (\varepsilon<\varepsilon_{b}) $} \\
							K \left( \frac{\varepsilon}{\varepsilon_{b}} \right)^{-\beta}  & \text{$ (\varepsilon \geq \varepsilon_{b}) $}.
					 \end{cases}
\end{equation}
Here, $K$ is the normalization factor which is determined by Equation (1). 
For our calculations, we use the minimum energy of 300 eV (which is likely to come from the self-absorption process) 
and the maximum energy of 10$^{13}$ eV (which roughly corresponds to 
the most optimistic 
energy of accelerated electrons)
before considering
photon attenuation by pair creation. 
Note that the results in our study are insensitive to these values as long as $\alpha < 2$ and $\beta > 2$.
 
\subsection{The Pair-creation Cutoff}
\label{subsec:pair}
High-energy photons cannot escape from a merged shell 
because of
the pair-creation processes such as 
$\gamma \gamma \rightarrow e^+ e^-$ and $e \gamma \rightarrow e e^+e^-$. The most important 
process is $\gamma \gamma \rightarrow e^+ e^-$ under typical conditions of GRBs \citep{2004ApJ...613.1072R}. 
We calculate the pair-creation cutoff energy for high-energy photons by considering only $\gamma \gamma \rightarrow e^+ e^-$.

The optical depth $\tau_{\gamma\gamma}$ for this process at some energy $\varepsilon'$ can be calculated for a 
given photon spectrum. We assume the power-law spectrum 
in Section \ref{subsec:radiation} and calculate 
the optical depth in the comoving frame of the outflow
as follows 
\citep{1967PhRv..155.1404G, 1987ApJ...319..643L, 1987MNRAS.227..403S, 2001ApJ...555..540L, 
2006ApJ...650.1004B, Murase:2007ya}:
\begin{eqnarray} 
\label{eq:opticaldepth}
\tau_{\gamma \gamma} (\varepsilon') \simeq \xi (\beta ) n' ( \tilde \varepsilon') \sigma_T \Delta', \notag \\
\left[ \tilde \varepsilon' = \frac{(m_{e} c^2)^2}{\varepsilon'} \right],
\end{eqnarray}
where the number density of photons whose energies are larger than $\tilde \varepsilon'$ is given by 
\begin{equation}
\label{eq:density}
n'(\tilde \varepsilon') \simeq 
K 
\int_{\tilde \varepsilon'} 
\left( \frac{\varepsilon'}{\varepsilon_{b}'} \right)^{-\beta} d\varepsilon'.
\end{equation}
Here $\tilde \varepsilon'$ is the energy of a photon which interacts 
with the photon of energy $\varepsilon'$ at the pair-creation 
threshold, and $\sigma_{T}$ is the Thomson cross section. 
Note that Equation (\ref{eq:density}) is 
valid when $\tilde \varepsilon'$ satisfies 
$\tilde \varepsilon' > \varepsilon'_{b}$, and 
we may approximate the optical depth as
$\tau_{\gamma \gamma}(\varepsilon') 
\simeq \tau_{\gamma \gamma}(\tilde \varepsilon'_{b})$
for $\tilde \varepsilon' < \varepsilon'_{b}$
if the spectrum is typical $\alpha \sim 1$
(see for more discussions in Sections \ref{sec:spectrum} and \ref{sec:discussion}). 
The number density of photons in the comoving 
frame $n'$ relates with the number density in the source frame 
$n$ as $n'(\varepsilon') \simeq n(\varepsilon' \Gamma)/\Gamma$ 
because the isotropically distributed photons, whose energy is $\varepsilon'$ in the comoving frame, are blueshifted to 
$\varepsilon \simeq \Gamma \varepsilon'$ on average. 
$\xi (\beta)$ is the numerical factor that depends on the power-law index and this factor decreases with $\beta$; the values 
are $\xi (\beta) =11/90\simeq 0.12$ and $\xi (\beta) = 7/75 \simeq 0.093$ for $\beta = 2$ and $\beta=3$, respectively 
\citep{1987MNRAS.227..403S}. We can use $\xi (\beta) \simeq 7 (\beta -1)/\left[ 6 \beta^{5/3} (\beta+1) \right]$ for 
$1 < \beta < 7$ for isotropically distributed photons \citep{2006ApJ...650.1004B}. 
We approximate the width of the emission region $\Delta'$ 
by using the width of the merged shell, that is, $\Delta' \simeq l'_{m}$
(see the Appendix A). 
The pair-creation cutoff energy $\varepsilon'_{\rm{cut}}$ is defined as the energy at which the optical depth becomes unity, 
$\tau_{\gamma \gamma} (\varepsilon'_{\rm{cut}} ) = 1$. 

For sufficiently small emission radii or Lorentz factors, the Thomson optical depth 
$\tau_T$ may exceed unity. 
Even in such cases, we always expect that the pair-creation cutoff should be comparable to $\sim \Gamma m_e c^2/(1+z)$. 
However, the Compton scattering is very 
important when $\tau_T \gtrsim 1$ 
\citep[see, e.g.,][]{1994MNRAS.270..480T,2004ApJ...613..448P,2005ApJ...628..847R}. 
The photon spectrum may have a thermal or 
quasi-thermal component \citep[see, e.g.,][]{2004ApJ...613..448P, 2006ApJ...642..995P}. It may deviate from 
the power-law shape and even have the pair annihilation bump \citep{2007ApJ...670L..77I, Murase:2007ya}.  
In this work, we treat the cutoff as $\varepsilon _{\rm{cut}} = \varepsilon_{\rm{ann}}=\Gamma m_e c^2/(1+z)$ for $\tau_{\gamma \gamma}(\Gamma m_e c^2/(1+z)) > 1$ 
for simplicity. Although this may not be a good approximation, we expect that this treatment is enough for our purpose since 
we do not have so many shell collisions with $\tau_{\gamma \gamma}(\Gamma m_e c^2/(1+z)) >1$ for our adopted parameter sets.

For sufficiently large emission radii or Lorentz factors, 
both the Thomson optical depth and the pair-creation 
optical depth may be very small, 
where no pair-creation cutoff exists because of $\tau_{\gamma \gamma}
({\varepsilon}^{\prime}) < 1$
for any energy ${\varepsilon}^{\prime}$.
Then, a significant fraction of $\gtrsim$ TeV photons escapes from the source 
without internal attenuation. These photons should be attenuated via pair creation by the cosmic microwave background (CMB) 
and cosmic infrared background (CIB) photons \citep[see, e.g.,][and references therein]{metal07}. 
We include this effect by using the "low-IR" model of \citet{2004A&A...413..807K}. 
Note that recent observations of TeV blazars imply that the low-IR model is favored \citep{2006Natur.440.1018A, 2007A&A...475L...9A}. 
The created pairs can affect the primary spectrum itself via the inverse Compton scattering of the CMB photons \citep{dl02,2004ApJ...613.1072R,metal07}. 
However, this effect leads to relatively minor effects during the prompt emission 
and can be important in the late time only when 
the intergalactic magnetic field is weak enough 
\citep{metal08,Takahashi:2008pc,2009MNRAS.396.1825M}.
Hence, we neglect this possible sub-dominant contribution from the pair echo emission in this work.   

\subsection{Parameters for Shells}

We have the following parameters in our calculations:
the initial length between two shells $d$, 
the initial width of a shell $l$, 
the number of shells $N$, 
the distance from GRB to the Earth $D$ (or the redshift $z$), 
the fraction of energy that goes to non-thermal electrons $\epsilon_{e}$, 
the initial number density of protons in the innermost shells $n_p$,
and the most inner radius of sub-shells $r_{\rm{ini}}$
(see also the Appendix B). 
We adjust the number density $n_p$ so as to have the given isotropic luminosity
$L_{\gamma}$.
In addition, we assume the log-normal distribution of 
shell's bulk Lorentz factors $\Gamma$, whose average value is $\Gamma_0$, 
with the fluctuation $A$. Note that $\Gamma_0$ is different from the mathematically strict mean value 
(see the Appendix B for definition of the log-normal distribution). 
We choose the log-normal distribution for a demonstrative purpose, although there is no observational 
evidence to support this distribution. However, one can see that the result does not change so much 
even if we use some other distributions (e.g., homogeneous distributions). 
Note that our aim here is not to reproduce observational relations such as a correlation between the peak 
energy in a spectrum and the isotropic luminosity in a burst, called the $\varepsilon_b$-$L_{\gamma}$ relation \citep{2004ApJ...606L..29L, 2004ApJ...609..935Y}. 
Examining distributions of parameters, taking into account such observational relations, would be important 
because it may include the information of the Lorentz factor distribution. 
But, such studies are beyond the scope of this work so that we just assume the log-normal distribution.
In this paper, the $\Lambda$CDM cosmology
with $\Omega_{m}=0.3$, $\Omega_{\Lambda}=0.7$, 
and $H_0 = 70$ km s$^{-1}$ Mpc$^{-1}$ 
is also adopted. 

\section{Results}
\label{sec:spectrum}
\begin{figure}[tb]
\centering
\includegraphics[height=7cm,angle=270,clip]{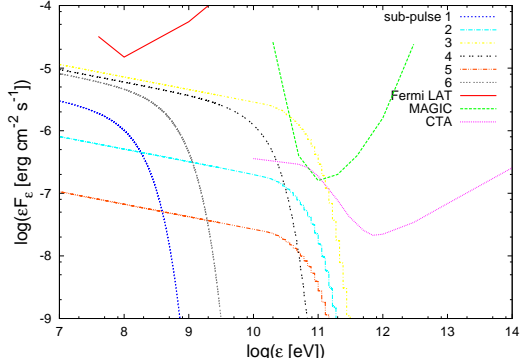}
\caption{$\varepsilon F_{\varepsilon} $ spectrum of prompt emission from each two-shell collision
where six spectra are randomly selected from dozens.
The number of shells is set to $N=100$ and 
the distribution of initial Lorentz factors is log-normal with $\Gamma_0 = 500$ 
and an initial fluctuation $A=1$, which are defined 
by Equation (\ref{eq:lognormal}). 
Other parameters are 
the initial distance between shells
$d=5\times 10^9$ cm, 
the initial thickness of a shell
$l=10^9$ cm, 
the initial radius of the innermost shell $r_{\rm ini}=10^{11}$ cm,
the energy fraction of electrons $\epsilon_{e} = 1/3$,
and the redshift $z=1$.
We adjust the initial number density $n_p$ 
so as to have the mean 
luminosity of $L_{\gamma}=10^{52}$ erg s$^{-1}$. 
The sensitivity curves of LAT on \textit{Fermi}, MAGIC, and CTA are also shown.
} 
\label{fig:manypulse}
\end{figure}
\begin{figure}[tb]
\centering
\includegraphics[height=7cm,angle=270,clip]{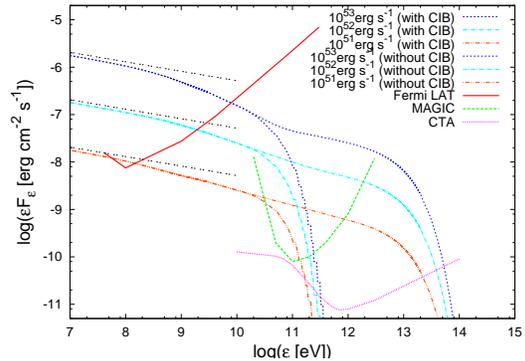}
\caption{Time-integrated spectra for different luminosities. 
We calculate three cases with almost the same 
pair-break energy (defined in Section \ref{sec:spectrum}) 
but different luminosities 
$L_{\gamma} = 10^{53}$, $10^{52}$, and $10^{51}$ erg s$^{-1}$. 
The fluctuation of the initial Lorentz factor $A$ is 0.5 
for $L_{\gamma} = 10^{53}$ erg s$^{-1}$, 
0.7 for 10$^{52}$ erg s$^{-1}$, and 1.6 for $10^{51}$ erg s$^{-1}$, respectively. 
Other parameters are the same as Figure \ref{fig:manypulse} except for 
the number density $n_p$. 
The number density is determined to set the luminosity to the above values. 
The slope of $\varepsilon^{-0.2}$ is shown with the dotted line for 
comparison. The sensitivity curves of LAT on \textit{Fermi}, MAGIC, and CTA are also shown.}
\label{fig:L53-51}
\end{figure}

In Figure \ref{fig:manypulse}, we show the energy spectra of emission 
from each two-shell collision using the method in Section \ref{sec:radiation}.
We only show six spectra randomly selected from a GRB
to make the figure easy to see, 
although there are dozens of spectra for one realization. 
We can see the electron$-$positron pair-creation cutoff in each spectrum. 
One may expect to obtain much information about 
the GRB if we can observe the cutoff energy. 
However, we have to integrate these spectra unless we can observe these spectra 
separately due to the lack of photon numbers. 
In Figure \ref{fig:manypulse}, all the spectra are below 
the sensitivity curve of \textit{Fermi} 
( the cutoff energy may be observed by MAGIC and/or CTA. But note that these observatories take a few 
dozens of seconds to turn the detector to the direction of a GRB after an alert from a space telescope). 
In this case, only the integrated spectrum is observable and 
the cutoff is smeared out as we explain below. 

\begin{figure*}[tb]
\centering
\includegraphics[height=7cm,angle=270,clip]{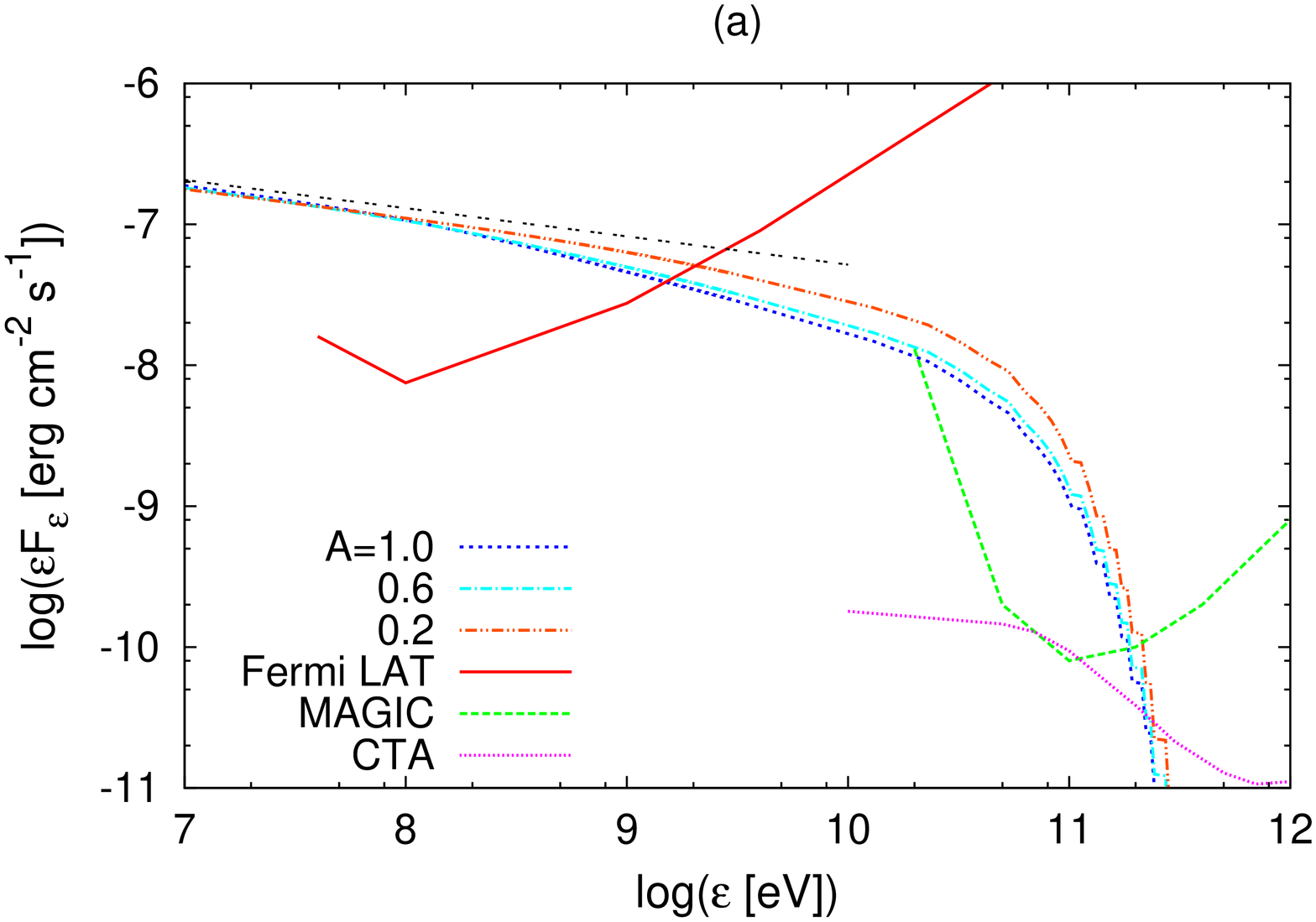}
\includegraphics[height=7cm,angle=270,clip]{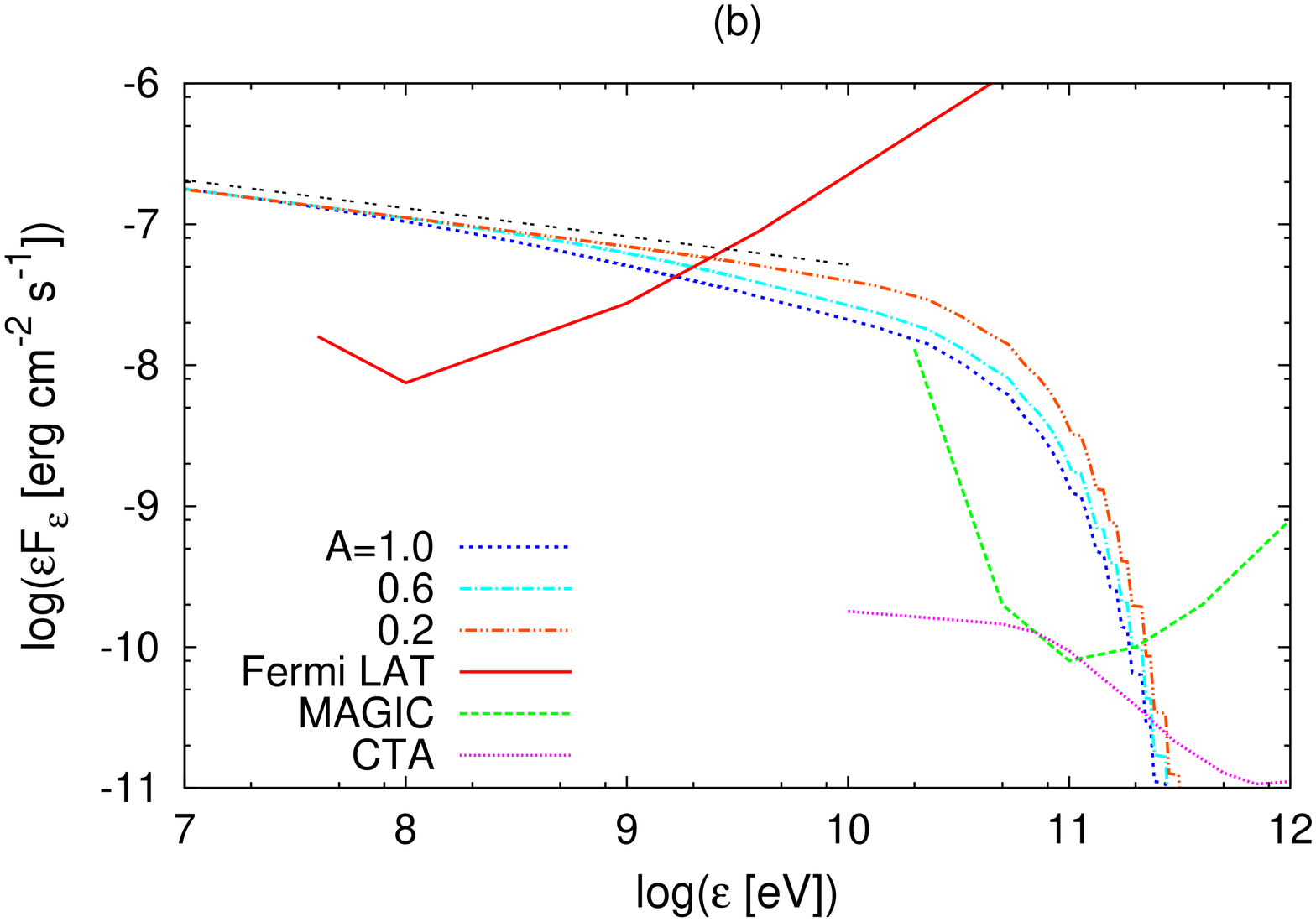}
\includegraphics[height=7cm,angle=270,clip]{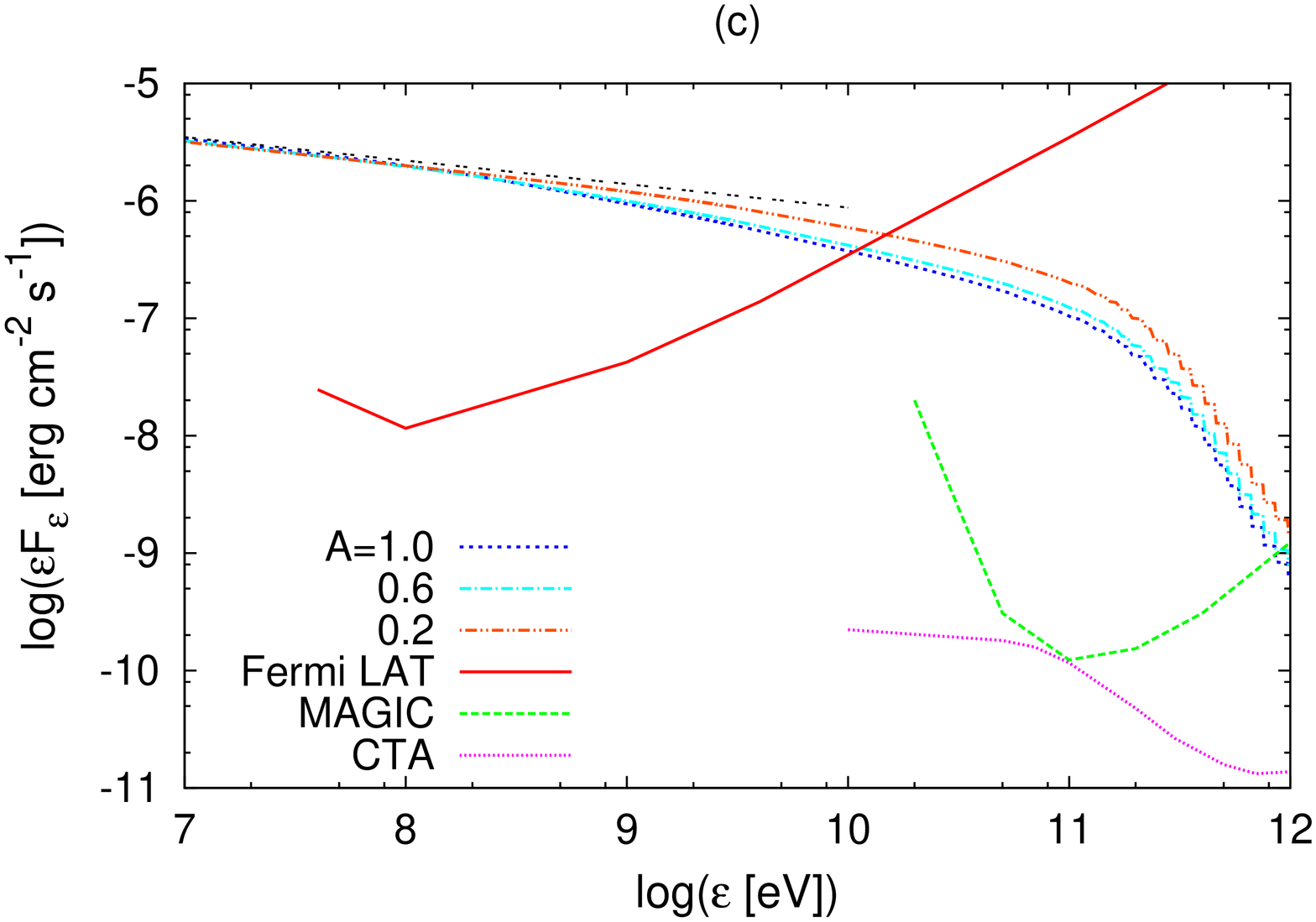}
\includegraphics[height=7cm,angle=270,clip]{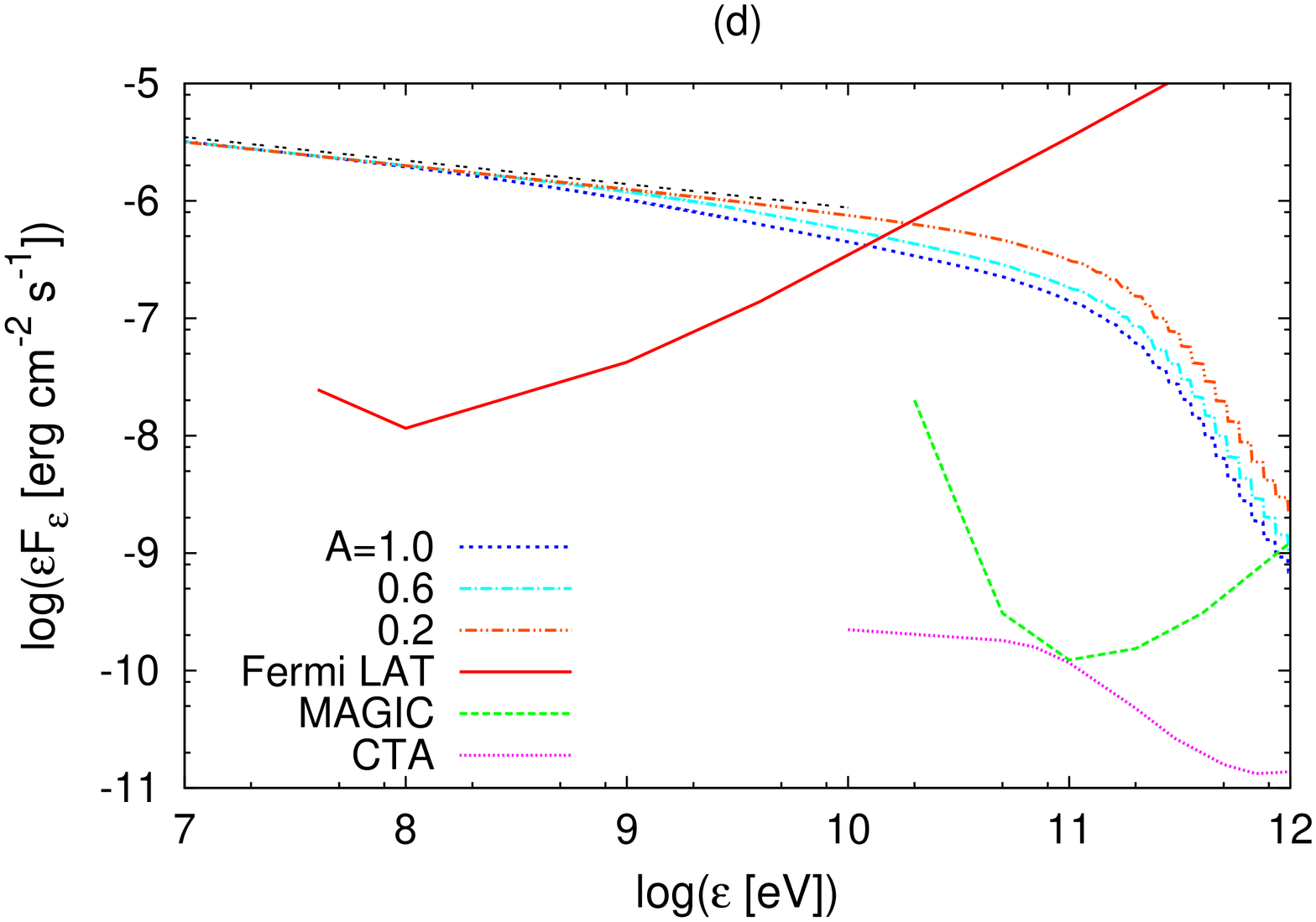}
\caption{Time-integrated spectra for the different fluctuation of Lorentz factor $A=1, 0.6, 0.2$. 
The luminosity is set to $L_{\gamma}=10^{52}$ erg s$^{-1}$ in these figures. 
Other parameters are as follows: (a) $z=1$, $l=10^8$ cm, 
$N=10^3$; (b) $z=1$, $l=10^9$ cm, $N=10^2$; (c) $z=0.3$, $l=10^8$ cm, $N=10^3$; (d) $z=0.3$, $l=10^9$ cm, $N=10^2$. 
We have $\Gamma_0 =500$, $d=5 \times l$, $r_{\rm ini}=10^{11}$ cm, and 
$\epsilon_{\rm e} = 1/3$ for all cases. 
The dotted line shows the slope of $\varepsilon^{-0.2}$ for comparison.
We consider the effect of CIB in TeV energy range. 
The sensitivity curves of LAT on \textit{Fermi}, MAGIC, and CTA are also shown.}
\label{fig:main}
\end{figure*}

Here, the sensitivity curve of \textit{Fermi} is calculated 
from the effective area \footnote{Shown at http://www-glast.slac.stanford.edu/software/IS/glast\_ lat \_ performance.htm.} under the criterion that at least five photons 
are collected. The sensitivity curve of MAGIC is roughly estimated from the effective area for a zenith angle of $20^\circ $ with the criterion that at least 10 photons are 
collected, although the actual detectability requires careful analysis \citep{2007ApJ...667..358A}.  
The sensitivity of CTA is also estimated from the public curve
\citep{2009arXiv0912.3742W}. Although the actual one depends on details that
are not available, it is shown just for demonstration by raising the
public one by $\sqrt{50\verb|hr|/20(1+z)\verb|s|}$.

In Figure \ref{fig:L53-51}, the three time-integrated spectra are shown 
whose isotropic luminosities are 10$^{53}$ erg s$^{-1}$, 10$^{52}$ erg s$^{-1}$, and 10$^{51}$ erg s$^{-1}$. 
The fluctuation of the initial Lorentz factor $A$ is 0.5 for the case of 10$^{53}$ erg s$^{-1}$, 0.7 for 10$^{52}$ erg s$^{-1}$, and 1.6 for 10$^{51}$ erg s$^{-1}$, respectively. We choose the values of 
$A$ so as to set the spectrum steepenings at the same energy (see below). 
Other parameters are described in the caption of the figure. 
We plot the spectra with and without the CIB effect.

From Figure \ref{fig:L53-51}, we can find the following features.

(1) Most importantly, there is no sharp cutoff that originates 
from the pair production in Figure \ref{fig:L53-51}.
Instead, the spectral slope becomes steep above a few hundred MeV.
We define the beginning point of the steepening as the pair-break energy.
This is because the steepening arises from the superposition of
the pair creation cutoff in each collision 
that is plotted in Figure \ref{fig:manypulse}. 
The exponential cutoffs are smeared by the time integration
and result in a steepening of the power-law spectrum. Note that this feature is also observed 
with other distributions of the initial Lorentz factor. 
\textit{Fermi} may observe the pair-break energy, if very bright bursts occur
in the future. For identifying the steepening feature, it would also
be favorable to observe the spectral slope at very high energies with
Cherenkov telescopes such as MAGIC and CTA.

(2) The steepening effect is conspicuous for the small 
fluctuation of the initial Lorentz factor $A$
(corresponding to the high luminosity case $L_{\gamma}=10^{53}$ erg s$^{-1}$)
while it is not for large $A$.
This is because the released energy 
decreases more rapidly for smaller $A$ 
as the collision radius of the two shells becomes larger
and accordingly as the cutoff energy of a two-shell collision becomes higher.
We discuss this effect analytically in Section \ref{sec:discussion}. 
The steepening effect may be observed by \textit{Fermi} 
when the luminosity is larger than $\sim 10^{52}$ erg s$^{-1}$. 

(3) Although the steepening is small for large $A$ at the pair-break energy,
the slope gradually becomes softer at larger energy
(see also Figure \ref{fig:main}).
This is because the fluctuation of Lorentz factor evolves 
smaller with time as shells collide with each other, asymptotically
approaching the small $A$ case
(see Section \ref{sec:discussion} for more discussions).

(4) The slope of the spectrum becomes hard, returning back to 
the original high-energy power-law index $\beta$ around TeV energy.
This means that the spectrum is produced by a two-shell collision
without the pair creation cutoff, i.e., by low-density shells.
We can roughly estimate the hardening energy (or the transparent energy)
as $\sim \Gamma^2 (m_e c^2)^2/ (1+z) \varepsilon_{b}$ 
(see Equation (\ref{eq:opticaldepth}))
in the following way. 
We assume the photon density is decreasing monotonously 
above the break energy $\varepsilon_{b}$. 
Then, as the photon density becomes low 
with large collision radius and/or small internal energy, 
the low energy range of the spectrum below $\varepsilon_{b}$ 
becomes relevant for the pair creation as target photons. 
However the photon number is insensitive to the photon energy
below $\varepsilon_{b}$ for the typical low-energy power-law index $\alpha=1$,
and hence the optical depth to the pair creation 
becomes almost constant for the photon energy
above $\sim \Gamma^2 (m_e c^2)^2/ (1+z) \varepsilon_{b}$.
This is why the slope becomes hard above this energy. 
Unfortunately, it is difficult to observe the hardening effect  by the effect 
of CIB (see below).

(5) There is a cutoff around $10^{13}$ eV.
This is not the pair-creation cutoff as discussed above, but the cutoff due to 
the maximum energy of electrons that we put for convenience. 
Note that the maximum energy of electrons itself depends on the
details of acceleration mechanisms that are not in focus of this work.

(6) If we include the effect of pair-creation interactions with CIB, 
we have a cutoff around TeV energy.
This effect masks the hardening effect and the maximum energy cutoff
discussed above.
All those features can be studied by observations of Cherenkov telescopes.

In Figure \ref{fig:main}, we show the time-integrated spectra 
for different fluctuations of initial Lorentz factor $A$ from 1 to 0.2
with the same luminosity $L_{\gamma} = 10^{52}$ erg s$^{-1}$. 
We consider the effect of CIB in TeV energy range. 
The spectrum becomes steep above a pair-break energy.
 A pair-break energy is low for large $A$ 
because the collision radius is small and hence the cutoff energy is small
for each collision.
The steepening is observable for large $A$ even if a GRB occurs at high $z\sim 1$ 
(Figure 3(a) and (b)). 
We can observe the steepening for small $A$ if the GRB occurs at low $z$ 
(Figure 3(c) and (d)). 
We also compare the cases of $N=10^3$ ($l=10^8$cm) 
with that of $N=10^2$ ($l=10^9$cm). 
The shells collide with each other in 
the inner radius for $N=10^3$ than for $N=10^2$
because the separation is small. 
(The separation between the shells is small 
when the thickness of the shell is small. 
This is because we assume the separation and the thickness are comparable.) 
It leads to the smaller pair-creation cutoff in each collision
and the pair-break energy becomes smaller.

\begin{figure}[tb]
\centering
\includegraphics[height=7cm,angle=270,clip]{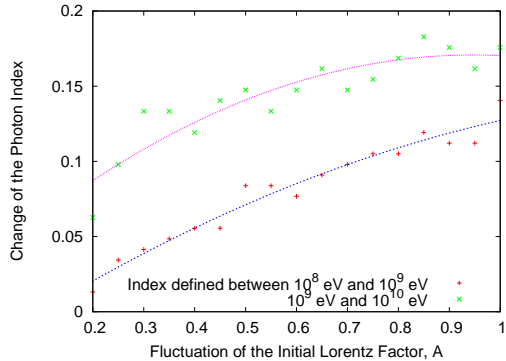}
\caption{Change of the photon index vs. the fluctuation of the initial Lorentz factor $A$. 
The change of the photon index indicates the difference from 
the initial high-energy power-law index $\beta$. 
The photon index is defined in
two different energy ranges, $10^8-10^9eV$ and $10^9-10^{10}eV$. 
The spectral index becomes larger as $A$ becomes larger. 
The solid and dashed lines are quadratic curves
obtained from the least-squares fitting.
}
\label{fig:index}
\end{figure}

In Figure \ref{fig:index}, 
we show a change of the power-law index from $\beta = 2.2$
due to the effect of the smearing for the case of Figure \ref{fig:main}(a) 
(the slope in Figure \ref{fig:main}(c) 
is the same as in Figure \ref{fig:main}(a)).
We calculate the smeared power-law index by two ways.
One is defined by the power-law index between $100$ MeV and $1$ GeV
with the least-squares fitting. 
This energy range corresponds to the observable energy band 
in Figure \ref{fig:main}(a). 
Another is defined by the energy range between 1 GeV and 10 GeV, 
which corresponds to the observable energy range in Figure \ref{fig:main}(c).
Figure \ref{fig:index} shows that the slope
in a fixed energy region is steeper for larger $A$.
The reason is that the pair-break energy is lower for larger $A$
(see the previous paragraph)
and the slope becomes softer as the fixed energy range is more separated from 
the pair-break energy (see the feature (3) obtained from Figure \ref{fig:L53-51}),
although the steepening is smaller at the pair-break energy
for larger $A$ (see the feature (2) obtained from Figure \ref{fig:L53-51}).
We have more discussions in Section \ref{sec:discussion}.

\begin{figure}[tb]
\centering
\includegraphics[height=7cm,angle=270,clip]{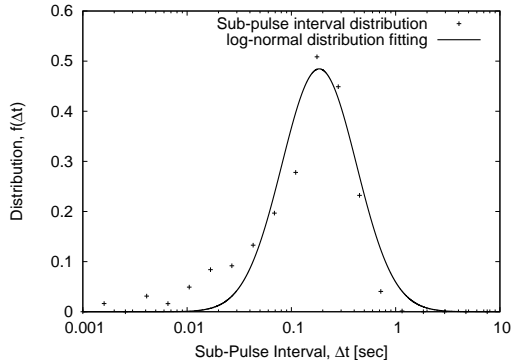}
\caption{Sub-pulse interval distribution. 
The parameters of the initial Lorentz factor distribution 
are $\Gamma_0 =  500 $ and $A=1$. Other parameters are the same as the 
calculation in Figure \ref{fig:main}(b). 
We fit the data of sub-pulse intervals by the log-normal distribution 
with the least-squares method.
}
\label{fig:pulse_interval}
\end{figure}
\begin{figure}[tb]
\centering
\includegraphics[height=7cm,angle=270,clip]{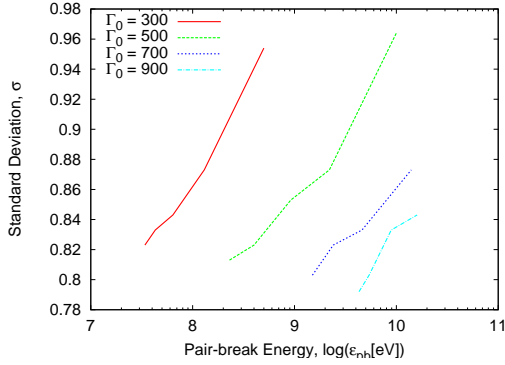}
\caption{Pair-break energy  vs. standard deviation 
of the sub-pulse interval distribution. 
The pair-break energy is defined by the energy 
where the spectrum becomes steeper than $\beta$ by 0.05. 
We calculate these values by changing $\Gamma_0$ and $A$, 
which are the parameters of the initial Lorentz factor distribution. 
The solid line is written for $\Gamma_0 = 300$ changing $A$ from 1 to 0.2.
The dashed line is written for $\Gamma_0 = 500$ changing $A$ from 1 to 0.2. 
The dotted line is written for $\Gamma_0 = 700$ changing $A$ from 1.3 to 0.4. 
The dot-dashed line is written for $\Gamma_0 = 900$ changing $A$ from 1.5 to 0.6. 
Other parameters are the same as the calculation in Figure \ref{fig:main}(b) 
}
\label{fig:smearing_sigma}
\end{figure}

We also calculate sub-pulse intervals. 
It is known that the distribution of sub-pulse intervals can be fitted 
by a log-normal distribution, although there is a small excess at long intervals 
\citep{1994MNRAS.271..662M,1996ApJ...469L.115L,2002MNRAS.331...40N,
2002ApJ...570L..21I}. 
We calculate sub-pulse intervals $\Delta t$ for 
various average Lorentz factors $\Gamma_0$ and fluctuation amplitudes $A$
(see the Appendix B for the definitions). 
We plot the result for $\Gamma_0=500$ and 
$A=1$ in Figure \ref{fig:pulse_interval}
as the histogram of the logarithmic time intervals.
The result is averaged for 100 different random realizations. 
A sub-pulse interval is defined by an interval between each beginning time of a sub-pulse. 
We fit the histogram by the log-normal distribution using 
the method of least squares. 
A log-normal distribution is characterized 
by $\mu$ and $\sigma ^2$ as $f (\Delta t ) d \ln \Delta t = 
\exp( - ( \ln(\Delta t ) - \mu )^2/2\sigma^2) /\sqrt{2\pi \sigma^2} d \ln 
\Delta t $.

In Figure \ref{fig:smearing_sigma}, we plot the correlation 
between the pair-break energy $\varepsilon_{\rm{pb}}$ and $\sigma$. 
We define the pair-break energy by the energy 
where the power-law index becomes steeper than $\beta$ by 0.05. 
We calculate $\varepsilon_{\rm{pb}}$ and $\sigma$ for various $\Gamma_0$ and $A$ 
fixing the luminosity. 
In this figure, $\varepsilon_{\rm{pb}}$ becomes smaller for larger $A$. 
The cutoff energy of two-shell collision is smaller 
when the fluctuation is large because the collision radius is smaller 
and the density of photons is larger. 
It means $\varepsilon_{\rm{pb}}$ becomes small when $A$ is large. 
It is difficult to observe $\varepsilon_{\rm{pb}}$ when the fluctuation is small 
by the effect of CIB. 
We discuss the possibility to extract the 
information from this plot in Section \ref{sec:discussion}.

\section{Discussion}
\label{sec:discussion}
\begin{figure}[tb]
\centering
\includegraphics[height=6cm,angle=0,clip]{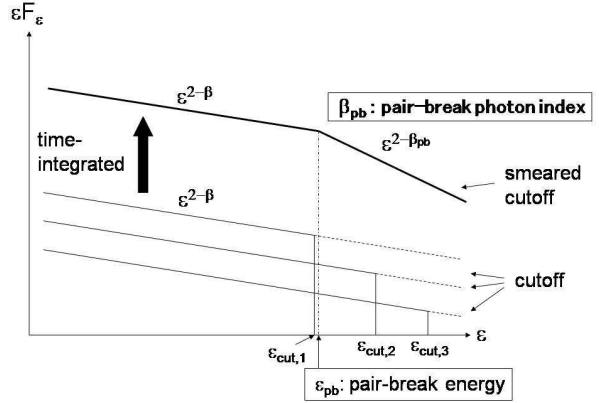}
\caption{Schematic picture of the GRB spectrum. 
The cutoff is smeared by the time integration, 
although each energy spectrum for the two-shell collision has a cutoff. 
There is no cutoff in the time-integrated spectrum 
but the slope becomes steep above the pair-break energy. 
This steep slope (the smeared cutoff) is made 
by the summation of different luminosity emissions. 
The pair-break energy is almost the same as the cutoff energy 
which originates from a two-shell collision at the most 
inner radius ($\varepsilon_{\rm{cut}1}$ in this picture).}
\label{fig:schematic}
\end{figure}
\subsection{On Estimate of Lorentz factors}
\label{subsec:lorentz factor}
Even though a pair-creation cutoff energy may give us useful information as we mentioned in Section \ref{sec:introduction}, 
it does not seem easy to obtain information when the cutoff is masked by the smearing effect (see Figure \ref{fig:schematic}). 
If our model assumptions are valid, it is difficult to see the obvious cutoff feature in the internal shock model.  
A clear cutoff feature should be in the very high energies of $\gg \rm
GeV$, but it is likely to be masked due to attenuation 
by the CIB. 
Then, can we probe the Lorentz factor from the GRB spectrum?
We think that we can still extract useful information on it.

One possibility is that we may constrain the Lorentz factor 
from the absence of the pair-creation cutoff. 
Since each cutoff is produced by the target photons
in each two-shell collision,
we may estimate each spectrum by dividing 
the time-integrated spectrum by the number of sub-pulses in the light curve.
However, this method usually overestimates the target photon density
because several sub-pulses may often overlap. In addition, observed
pulse durations and intervals may not be ideal sub-pulse durations (that are
directly calculated in the internal shock model) due to the overlap. 
In particular, high-energy gamma rays related to the maximum cutoff
come from the outer collision with a wide sub-pulse, which would be buried 
in many other spiky sub-pulses. 
Hence, we expect that the cutoff smearing effect can generally
reduce the conventional estimate of the Lorentz factor.

One may see the pair-break energy in high-energy spectra. 
The pair-break energy is almost the same as the minimum cutoff energy 
in two-shell collisions.
Hence, the pair-break energy would be useful instead of 
the original cutoff energy, if it is clearly observed. 
Although its observational identification may not be easy, 
let us consider the case that it is accomplished. 
Then, we can apply the conventional procedure described by various authors 
\citep[e.g.,][]{2001ApJ...555..540L,Murase:2007ya}. 
From Equations (\ref{eq:eint}) and (\ref{eq:density}), we have 
(for $\beta>2$)
\begin{eqnarray}
\epsilon_{e} E_{\rm int} &\simeq& V K (1+z) \frac{\varepsilon_{b}^2}{\beta-2},
\\
n'(\tilde \varepsilon') &\simeq& \frac{K}{\Gamma} 
\frac{(1+z) \varepsilon_{b}}{\beta-1} 
\left(\frac{\tilde \varepsilon}{\varepsilon_{b}}\right)^{1-\beta}.
\label{eq:density2}
\end{eqnarray}
By substituting these equations into Equation (\ref{eq:opticaldepth}),
we obtain the Lorentz factor as
\begin{eqnarray}
\Gamma  &\simeq&  \left( \frac{\varepsilon_{\rm{pb}} \varepsilon_{\rm b} (1+z)^2}{(m_e c^2)^2} \right)^{\frac{\beta-1}{2\beta +2}} \nonumber \\
& & \times \left( \frac{\xi \sigma_T \epsilon_{e} E_{\rm{int}} (1+z)^2 (\beta-2 ) }
{16 \pi c^2 \varepsilon_{b} \delta t^2 (\beta-1)} \right)^{\frac{1}{2\beta+2}},
\label{eq:gamma}
\end{eqnarray}
where we use $V \simeq 4 \pi r^2 \Delta$ and 
$r \sim 2 \Gamma^2 \Delta \sim 2 \Gamma^2 c \delta t/(1+z)$ 
expected in the internal shock model
(and we can also expect $r \simeq 2 \Gamma^2 c {\tilde{\delta} t}/(1+z)$ 
if an observed pulse duration $\tilde{\delta} t$ is the pulse decay time due to the 
curvature effect). 
For typical parameters, we have
\begin{eqnarray}
\Gamma &\simeq & 3.2 \times 10^2 \left( \frac{\varepsilon_{\rm{pb}}}{500\rm{MeV}} \right)^{\frac{3}{16}} (1+z)^{\frac{11}{16}} \left( \frac{\delta t}{0.1 \rm{s}} \right)^{-\frac{5}{16}} \nonumber \\
             & &  \times \left( \frac{\epsilon_{e} E_{\rm{int}}}{10^{53} \rm{erg }} \right)^{\frac{5}{32}}~~~~(\beta=2.2, \varepsilon_{b}=300\rm{keV} ),
\label{eq:gamma2}
\end{eqnarray}
where we note that $\epsilon_{e} E_{\rm int}$ 
is the released radiation energy per collision.

How could we estimate the Lorentz factor from the time-integrated spectrum? 
First, we could roughly estimate the released energy from 
each collision, $\epsilon_{e} E_{\rm int}$ 
by dividing the time-integrated spectrum by the sub-pulse number of the light curve, 
which may be observed in lower energy range near $\varepsilon_{b}$, 
once $D$ is determined by other observations. 
We could also estimate $\delta t / (1+z) \sim l/c \sim d/c \sim \Delta t$ if we can know the sub-pulse 
width in light curves, 
although it is difficult to evaluate these values accurately from observations. 
Then, we could roughly estimate the Lorentz factor $\Gamma$ 
from the pair-break energy $\varepsilon_{\rm{pb}}$. 
One advantage of using the pair-break energy is that
the minimum cutoff energy, i.e., the pair-break energy, 
is predominantly produced by the inner collision
with a short sub-pulse width, which may be directly identified with 
reasonable values of the observed shortest variability timescale.
Another advantage is that the pair-break energy is smaller than the cutoff energy in general and 
it can be smaller than the CIB attenuation energy. 
Several sub-pulses still often overlap, which may lead to our
overestimating the released energy from each collision $\epsilon_{e} 
E_{\rm int}$ via overestimating the duration and interval of sub-pulses 
from the most inner radii. Hence, as long as we can choose 
a reasonable shortest value of $\delta t$, this method could 
give an upper limit of the Lorentz factor (i.e., the true value of 
the Lorentz factor $\Gamma_{\rm{true}}$, which is obtained considering 
masked sub-pulses, is related with the Lorentz factor calculated by 
our formulation $\Gamma_{\rm{upper}}$ and a conventional 
estimate of the Lorentz factor $\Gamma_{\rm{co}}$ as follows: 
$\Gamma_{\rm{true}} \leq \Gamma_{\rm{upper}} \leq 
\Gamma_{\rm{co}}$). Again, note that conventional estimate generally lead
to overestimating the minimum Lorentz factor when we use short timescales of $\delta t$ since true sub-pulses related to the high-energy
emission may be wide and masked by other sub-pulses.
 
In the internal shock model, the Lorentz factor is likely to have a large dispersion, i.e., large $A$.    
Since the dispersion of the Lorentz factor $A$ is reflected by the dispersion of the sub-pulse interval $\sigma$, we 
may expect some correlation between $\sigma$ and $\varepsilon_{\rm{pb}}$ 
(see Figure \ref{fig:smearing_sigma}). 
Although this $\sigma$ is not the observed 
dispersion itself which depends on algorithms determining pulse widths (e.g., \citet{1996ApJ...469L.115L}), it 
suggests that we can obtain some potential clue to the average value of the Lorentz factor from 
high-energy gamma-ray observations.

\subsection{Comparison with Analytical Calculation}
\label{subsec:analytical}
We can calculate a power-law index produced by the smeared cutoff analytically 
if an initial variance of the sub-shells' Lorentz factor distribution $\sigma^2_{\Gamma,0}$ satisfies 
$\sigma^2_{\Gamma, 0} < \Gamma_0^2$. 
Here $\Gamma_0$ is an average Lorentz factor and 
$\sigma^2_{\Gamma, 0}$ is defined in the source frame. 
Other assumptions are the same as the numerical calculation in this study. 
In this case, the variance of the shells' velocity distribution $\sigma_v$, 
which is defined in the center of mass, evolves with a radius of the expanding 
shell $r$ as $\sigma_v \propto r^{-1/3}$. 
The internal energy that originates from the two-shell 
collision is proportional to $\sigma_v^2$ \citep{2000ApJ...539L..25B} and 
this energy decreases as $E_{\rm{int}} \propto \sigma_v^2 \propto r^{-2/3} $ with radius 
\citep{2008ApJ...674L..65L}. The internal energy is converted to non-thermal energy of electrons 
that radiate photons. 
We discuss a situation that a photon 
whose energy is $\varepsilon_{\rm{cut}} $ annihilates with a photon whose energy is 
$\tilde \varepsilon ~(> \varepsilon_b)$. Here, $\tilde \varepsilon$ 
is the typical energy of target photons. 
Then the target photon number density is written as 
$n \propto E_{\rm{int}} \tilde \varepsilon^{1-\beta} /4 \pi r^2 \Delta 
\Gamma^2 \propto r^{-8/3} \tilde \varepsilon^{1-\beta} \Delta^{-1}$. 
We can calculate how the optical depth depends on $r$ and 
$\tilde \varepsilon$, 
$\tau_{\gamma \gamma} (\varepsilon )\propto 
n \sigma_T \Delta \propto \tilde \varepsilon^{1-\beta} r^{-8/3}$.
The target photon energy scales as $\tilde \varepsilon \propto r^{8/3(1-\beta)} $. High-energy photons with energy $\varepsilon$ that 
satisfies $\varepsilon \geq (\Gamma m_{e} c^2)^2/(1+z) \tilde \varepsilon$ produce pairs in interaction with photons. We evaluate $\varepsilon_{\rm{cut}}$ 
using the minimum energy for the pair-creation threshold. Then, the cutoff energy scales as 
$\varepsilon_{\rm{cut}} \propto r^{-8/3(1-\beta)}.$ As we explain above, the internal energy which associates 
with the fluctuation of shell velocities decreases with shell's radius as $E_{\rm{int}} \propto r^{-2/3}$. Hence 
$E_{\rm{int}} $ relates with the cutoff energy as $E_{\rm{int}} \propto \varepsilon_{\rm{cut}}^{(1-\beta)/4}$. 
The energy flux at $\varepsilon_{\rm{cut}}$ is in proportion to $E_{\rm{int}}$ and $\varepsilon_{\rm{cut}}^{2-\beta}$ if we assume the energy spectrum that originates from 
two-shell collision 
satisfies the power law and the break energy (or peak energy) of each shell's spectrum does not change. The high-energy part of time-integrated spectrum would 
be 
\begin{eqnarray}
\varepsilon F_\varepsilon (\varepsilon) \propto E_{\rm{int}} \varepsilon^{2-\beta} \propto \varepsilon^{\frac{9-5\beta}{4}}.
\end{eqnarray}
The pair-break photon index $\beta_{\rm{pb}}$ (see Figure \ref{fig:schematic}) becomes $(5\beta-1)/4$ and the change of the index is $(\beta-1)/4$
($= 0.3$ for $\beta=2.2$).

For large fluctuation of the Lorentz factor 
$\sigma_{\Gamma,0}^2 > \Gamma_0^2$, i.e., $A>1$,
we cannot apply the above analytical results.
Actually we have shown that the steepening is more mild
for larger $A$ in Section \ref{sec:spectrum}.

In Figure \ref{fig:main}, we plot the time-integrated spectrum. The beginning energy of steepening 
depends on the fluctuation of the initial Lorentz factor $A$. The 
steepening begins at low energy when $A$ is large. The collision radius is small for large $A$ and 
the density of photons becomes 
large. This leads to the small cutoff energy of the two-shell collision and 
the steepening begins at lower energy. 
For large $A$, the steepening is small at the pair-break energy
(see Section \ref{sec:spectrum}), but
the slope becomes steep gradually at high energy because the fluctuation 
of Lorentz factor becomes small as shells collide with each other and the slope approaches the behavior of the analytical solution. In fact, $E_{\rm{int}}$ follows the analytical 
solution, $\propto r^{-2/3}$, after some collisions. However, there is little difference between the numerical calculation and the analytical one even if the slope becomes 
steep. This can be understood as follows (see also Section \ref{sec:spectrum}). 
Photons whose energy is lower than $\varepsilon_{b}$, which is defined in Section \ref{subsec:radiation}, is important when the 
cutoff energy is large. The number of these photons 
are small to interact with other photons and the cutoff energy becomes large suddenly. In our numerical calculation, we approximate that there is no cutoff when the cutoff 
takes such a large value. On the other hand, 
we consider a single power law without the spectrum break at 
$\varepsilon_{b}$ in the analytical calculation. 
In this case, there are many low-energy photons and the cutoff energy increases following the power law of the collision radius. We checked that the numerical calculation gives 
the same result as the analytical calculation in a low $A$ regime when we assume the photon spectrum satisfies a single power law for a test numerical calculation. We 
can understand the behavior of the 
smeared cutoff in Figure \ref{fig:L53-51} similarly. The pair-break photon index which is defined between 1GeV and 10GeV is 0.4 for $L_{\gamma}=10^{53}$ erg s$^{-1}$, 
0.3 for $L_{\gamma} = 10^{52}$ erg s$^{-1}$, and 0.3 for $L_{\gamma} = 10^{51}$ erg s$^{-1}$, respectively. 

\subsection{Comparison with Recent Observations of \textit{Fermi}}
We can estimate the Lorentz factor of a GRB 
if we observe a cutoff in a single sub-pulse spectrum. 
Then one may think that the maximum energy of the observed photon 
gives a lower limit of the Lorentz factor  
\citep{1997ApJ...491..663B,2001ApJ...555..540L,2004ApJ...613.1072R}. 
However, as we have discussed in Section \ref{subsec:lorentz factor},
the cutoff smearing effect
generally allows lower Lorentz factor than the conventional estimates.

In the case of GRB080916C, $3$ GeV photon was detected by \textit{Fermi} Large Area Telescope (LAT) in first 10 s and $13$GeV photon was detected in 100 s 
\citep{2009Science...080916}. 
In \cite{2009Science...080916}, there are five bins ($a-e$) in 
the light curve. The spectrum index is almost constant from time interval 
$b$ to $e$. Hence, we can interpret 
the observation of beta as constant in one time interval of the observation, and 
it seems to be reasonable to interpret like this. We use the time interval $b$ to estimate 
the Lorentz factor in a following discussion. 
We can constrain the Lorentz factor using the maximum energy of a detected photon. 
The minimum Lorentz factor is 
 $890$ when we use following parameters, 
$\varepsilon_{b}=1.17\times10^6\rm{eV}$, 
$\varepsilon =3 \times10^9\rm{eV}$, $\beta=2.21$, $z=4.35$, 
$E_{\rm{int}}=4.2 \times 10^{54}$erg and $\delta t=2\rm{s}$. 
Here,  the observed time interval $\tilde{\delta} t=2\rm{s}$ is regarded as  the single sub-pulse width $\delta t$ with energy $E_{\rm{int}}$. 
However, there may be a pair-break energy because 
the spectral steepening (by less than $(\beta-1)/4 \sim 0.3$)
is too small to be detected with
the current high energy data.
When we set the sub-pulse duration to the high temporal variability, $\delta t \sim 100$~ms 
detected by International Gamma-Ray Astrophysics Laboratory \citep[INTEGRAL;][]{2009A&A...498...89G} 
and $E_{\rm{int}}=4 \times 10^{52}$ erg accordingly,
the conventional estimate on the minimum Lorentz factor gives $\sim
1100$.
However, this value may be overestimated. Here, let us apply our
estimate using the pair-break energy. Then, we can obtain 
the Lorentz factor as $\sim 600  {(\varepsilon_{\rm pb}/100 \rm
MeV)}^{3/16}$ for $E_{\rm{int}}=7 \times 10^{52}$ erg and $\delta
t=100$ ms.
The actual Lorentz factor could be even lower as long as 
our choice of $\delta t$ is reasonable, 
because the spectrum consists of many sub-pulses in the cutoff smearing picture 
leading to overestimating the single sub-pulse energy $E_{\rm{int}}$
(see Equations (\ref{eq:gamma}) and (\ref{eq:gamma2})).

A hard lag is observed in the observation of GRB080916C. 
It might suggest that the time development of a cutoff energy. 
Generally, the cutoff energy from an inner radius is smaller than that from an outer radius. The hard lag might be explained by multiple collision of 
shells because a photon from a inner radius seems to be detected first. 
However, we did not find the time lag in the different energy range in our 
calculation. A photon from an inner radius can be detected 
after a photon from an outer radius because the shell is distributed with 
a moderate width. 
It does not seem so easy to reproduce the time lag in the current model.

\subsection{Effects of Radiation and Acceleration Mechanisms}
\label{subsec:other}
In this study, we do not consider an additional high-energy component in the prompt emission spectrum. 
Such an additional component has been often expected in the optically thin synchrotron scenario \citep[e.g.,][]{2006RPPh...69.2259M,2007ChJAA...7....1Z}. 
Theoretically, there are two main classes as high-energy emission mechanisms, i.e., leptonic and hadronic mechanisms.
The leptonic mechanisms include synchrotron self-Compton (SSC) emission and external inverse-Compton emission,
which are the most discussed scenarios. High-energy SSC emission is produced by relativistic electrons that radiate 
seed synchrotron photons themselves \citep[e.g.,][]{pm96,dl02,gg03}. 
In addition, there may be some possibilities for external inverse-Compton emission via, e.g., upscattering thermal photons 
in the cocoon 
or non-thermal photons radiated from inner shells. In the later case, the integrated spectrum becomes 
softer than that of our model \citep{2010ApJ...709..525L}. 
The hadronic mechanisms include synchrotron radiation of high-energy baryons, synchrotron radiation
of the secondary leptons generated in photohadronic interactions, as well as the photons directly produced from 
$\pi^0$ decays. In order to see the baryon synchrotron radiation, sufficiently strong magnetic fields are typically required
\citep[e.g.,][]{gz07,metal08}.
Otherwise, photohadronic components would dominate over the 
baryon synchrotron component as long as the photon density is high enough 
\citep{da04,2007ApJ...671..645A}. 
Hadronic gamma rays can be observed only when the non-thermal baryon loading is large enough 
\citep[e.g.,][]{mn06, 2007ApJ...671..645A}. 
The above mechanisms can lead to an additional high-energy component that can be observed by \textit{Fermi}, and 
high-energy spectra may differ from those demonstrated in this work. Although they are potentially important, we postpone 
detailed discussions in this work for our demonstrative purpose.
  
Another uncertainty raises from the origin of the cutoff in high-energy spectra. For example, the maximum energy of generated 
photons is at most the maximum energy of electrons, which is often determined by comparing between the synchrotron cooling 
time and acceleration time in the optically thin synchrotron scenario. 
When the SSC emission is not significant, if the pair-creation opacity is small enough or the acceleration efficiency is small enough, the cutoff would be determined by the electron maximum energy rather than the 
pair-creation process \citep[e.g.,][]{2010arXiv1006.3073I}. 
In addition, 
other processes such as the Thomson scattering may also be relevant 
\citep[e.g.,][]{2004ApJ...613..448P}.   
Also, the acceleration mechanism affects the electron spectrum itself as well as the maximum energy of electrons. 
Generally speaking, the relativistic diffusive shock acceleration sensitively depends on properties of magnetic 
fields around shocks, so that the spectral index may not be universal and the spectrum may differ from the simple power law 
\citep[e.g.,][]{2001MNRAS.328..393A, 2005ApJ...626..877B, 2006MNRAS.366..635L, 2006ApJ...641..984N,   2007ApJ...658.1069M, 2007ApJ...662..980M, 2008MNRAS.383.1431A}. 
These effects from radiation and acceleration mechanisms may significantly affect our results of high-energy spectra made by 
the smeared pair-creation cutoff.

\subsection{Effects of radiative transfer}
\label{subsec:radiative}
\begin{figure}[tb]
\includegraphics[height=8cm,angle=270,clip]{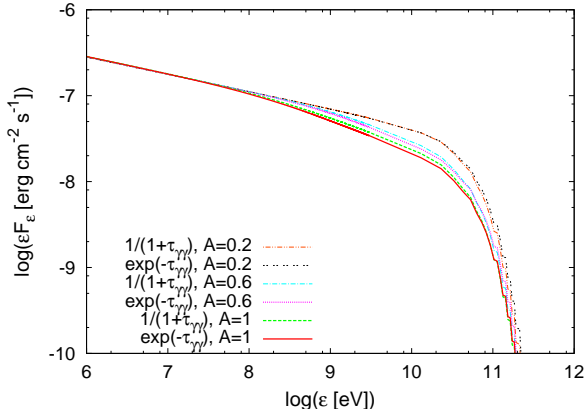}
\caption{Time-integrated spectra for two models. Curves labeled as $\exp(-\tau_{\gamma \gamma})$ 
represent the spectrum that is summed over the spectra with the exponential cutoff. Curves labeled as 
$1/(1+\tau_{\gamma \gamma})$ 
represent the spectrum that is summed over the spectra with  the $1/(1+\tau_{\gamma \gamma})$ attenuation signature. Curves show similar profile. 
Parameters are the same as the calculation in Figure \ref{fig:main}(b). 
}
\label{fig:compare}
\end{figure}

In this study, we calculated the energy spectrum assuming that each shell is regarded as a time-independent 
one-zone shell. 
One may think about effects of spatially distributed photons and/or time-dependent formulations. 
However, our conclusion does not change so much even if we consider these effects. We describe explanations 
below.
First, let us consider the effect of the spatially distributed photons. High-energy photons are 
preferentially weighted by contributions of outer regions because photons of outer regions 
typically have the lower optical depth for pair-creation processes. One obtains  a broken power-law 
energy spectrum even for emission from a single shell, which is different from the spectrum with the 
exponential cutoff (see Section \ref{subsec:pair}). 
However, we can argue that the result in the overall spectrum does not change dramatically even if we use the exponential 
cutoff. Let us examine the case of $1/(1+\tau_{\gamma \gamma})$ attenuation signature, which is valid 
when we consider 
the effect of spatially distributed photons \citep[e.g.,][]{2006ApJ...650.1004B}. $\tau_{\gamma \gamma}$ 
is in proportion to $\varepsilon^{\beta-1}$ in our model (see Equations \ref{eq:opticaldepth} and 
\ref{eq:density2}). Thus, a slope of a single-shell energy spectrum changes from $-\beta$ to $-2\beta+1$ 
at $\varepsilon_{\rm{cut}}$. 
As a result of our calculation, the slope of the time-integrated spectrum changes from $-\beta$ to $-(5\beta-1)/4$ 
by the effect of multiple shells even if fluctuation of the initial Lorentz factor distribution $A$ is small  (see 
Section \ref{subsec:analytical}). The 
change of the slope becomes smaller if $A$ is larger. Hence the consequence of $1/(1+\tau_{\gamma \gamma})$ attenuation 
signature is masked by the effect of multiple shells as far as $\beta > 1$, because the change of slope due to 
$1/(1+\tau_{\gamma \gamma})$ attenuation is larger than that due to the effect of multiple shells. 
This means that the exponential cutoff and 
$1/(1+\tau_{\gamma \gamma})$ attenuation gives similar results, and our conclusion does not depend 
on the choice.
In fact, the exponential cutoff and $1/(1+\tau_{\gamma \gamma})$ give similar spectra in Figure 
\ref{fig:compare}. 
Note that the exponential turnovers in this figure come from the attenuation by the CIB. 
Two spectra are almost same for small $A$. On the other hand, there is small difference in a high-energy range when the above analytical solution is not good (i.e., large $A$). The spectrum 
with $1/(1+\tau_{\gamma \gamma})$ attenuation signature is larger than that with the exponential 
cutoff. This is because emission from inner radius is large for large $A$ and a 
$1/(1+\tau_{\gamma \gamma})$ tail is comparable to a spectrum of emission from outer radius. 
Another effect of spatially distributed photons also arises if electrons are in the fast cooling range and 
particles are accelerated by the diffusive shock acceleration, since such electrons are expected to be 
confined to the emission region close to the shock. The localization of the fast cooling electrons may 
affect our results (Granot et al. 2008).

Next, we examine the effect of a time-dependent formulation. There is an intermediate power-law segment in the 
time integrated spectrum when we consider the time development of the photon field and the geometrical 
effect, under the assumption that a thin shell expands and emits isotropically in its own rest frame 
\citep{2008ApJ...677...92G}. The slope of the spectrum is $-\beta$, $1-2\beta$ and $-(\beta-1)(2\beta+3)/
(\beta+1)-1$, respectively, for impulsive sources (i.e., for thin shells). The slope of the intermediate 
power law is steeper than that of the spectrum integrated over multiple shells. 
Then, the time-integrated 
spectrum resulting from our calculation would not be changed as well as we discussed in the last paragraph 
for the case of the spatially distributed photons, if their formulation is used. 
This is because a change of the slope by the time-dependent effect is steeper 
than that by the effect of multiple shells. Note that 
the first break energy \citep[labeled as $\varepsilon_{1i}$ in][]{2008ApJ...677...92G} 
with the time-dependent calculation is equal to $\varepsilon_{\rm{cut}}$ determined from time-independent 
one-zone calculation. The difference between the two formulations is masked once we 
integrate spectra of photons emitted from each shell, so that our conclusion does not change so much by 
this effect  as long as we suppose impulsive sources.

Note that the above discussions are correct under our assumption (see Section \ref{sec:radiation}). The effects of  the radiative transfer may not be remarkable unless photons are not emitted from the outer region effectively (i.e., the effect of the spatially distributed photons and/or the time-dependent are weaker than that of the multiple shells). For example, an amount of high-energy photons becomes less if $\beta$ and $\varepsilon_{b}$ 
decrease as a collision radius becomes large.

\begin{figure}[tb]
\includegraphics[height=8cm,angle=270,clip]{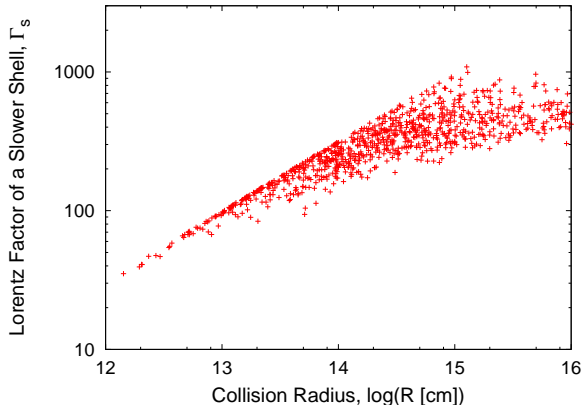}
\caption{Collision radius vs. a Lorentz factor of a slower shell. These two values are correlated with 
each other when an initial fluctuation of Lorentz factor $A$ is large. We calculated these values for 
$A=1$. Other parameters are the same as the calculation in Figure \ref{fig:main}(a). }
\label{fig:correlation}
\end{figure}

Finally, let us discuss the effect of the opacity by outer shells. Photons emitted from inner shells are absorbed in 
outer shells if $\varepsilon_{\rm{cut}}$ of photons emitted from inner shells are larger than that from outer shells. 
However, $\varepsilon_{\rm{cut}}$ is small at inner shells if shells' Lorentz factors are small in an inner radius. 
This trend is obvious for large $A$ because a Lorentz factor distribution of shells becomes more important 
than a spatial distribution of shells to determine a collision radius. 
In other words, a shell whose Lorentz factor is sufficiently small collides with another shell in an inner 
region even if an initial position of this shell is relatively outer region compared with other shells. 
Note that the collision radius is 
determined by a velocity of a caught-up shell (i.e., a slower shell) and the distribution of collision radii is 
correlated with the distribution of slower shells (Figure \ref{fig:correlation}). On the other hand, a correlation 
between the Lorentz factor 
and a collision radius is weak for small $A$ because the effect of the spatial distribution is remarkable. 
Hence, when $A$ is large, outer sub-shells do not contribute to the pair-creation opacity for photons emitted from inner sub-shells unless energy of radiated photons from outer radii is rather larger than that from inner radii. Otherwise, photons from inner sub-shells suffer from $e^{-\tau_{\gamma \gamma}}$ in outer sub-shells. 

\section{Summary and Conclusion}
We have investigated the spectral features
of the electron$-$positron pair creation
in the GRB prompt emission
to probe the Lorentz factor of the GRB outflow.
We have studied the time-integrated spectrum
from many internal shocks
with numerical and analytical calculations.
This is because almost all the authors studied the pair-creation cutoff energy expected from one emission region, although 
it could be difficult to observe the emission from a single collision. 
Although the cutoff energy may be observed for a simple 
sub-pulse from bright bursts and it is very useful 
\citep[e.g.,][]{Murase:2007ya,2008MNRAS.384L..11G}, it is also important to 
study whether the cutoff can be observed or not for bursts that have many sub-pulses in their light curves. 
\textit{Fermi} typically allows us to obtain the high-energy spectrum integrated over some time interval (e.g., $\tilde{\delta}=2$~s) in many cases due to the difficulty to collect many high-energy photons and the overlapping of sub-pulses. 
In this sense, current and future Cherenkov detectors with the low-energy threshold of a few $\times 10$~GeV are important to study the cutoff because these observatories can collect many high-energy photons. 
MAGIC and VERITAS should be useful for nearby GRBs, and future CTA will allow us to have more chances to observe GRBs with high-energy emission.    

We demonstrate that
the clear cutoff, which comes from electron$-$positron pair-creation processes, 
is hard to observe 
in the internal shock model, when bursts consist of many sub-shell
collisions. 
Instead, the slope of the spectrum becomes steeper above the energy, 
so-called the 
pair-break energy $\varepsilon_{\rm{pb}}$, 
which 
also originates from the electron$-$positron pair creation. 
We have demonstrated how the cutoff is smeared, and compared our numerical results with analytical consideration.
Possibly, we expect that \textit{Fermi} observes this pair-break energy in the high-energy range.
The pair-break energy may be observable if the luminosity is greater than $\sim 10^{52}$ erg s$^{-1}$ and 
the fluctuation of the initial Lorentz factor $A$ is large for $z=1$. 
However, the change of the spectrum slope is small in general and it may not be easy to observe the pair-break 
energy. 
Hence, observing nearby GRBs is important to identify the pair-break energy. 
It is also important to identify the pair-break energy by multi-wavelength observation using 
Cherenkov detectors. 
If it is detected, we have some information on the Lorentz factor of the shell emitting photons with $\sim \varepsilon_{\rm{pb}}$.
 
It was conventionally discussed that the pair-creation cutoff is
a broken power law (not an exponential one) since the observed gamma rays
and target photons are co-spatially and/or time-dependently distributed
in a single collision. However, our multiple shell effect would be more
prominent and mask the above effect with smaller change of the power-law
index as long as $\beta > 1$ and we suppose impulsive sources.

The cutoff smearing effect can generally reduce 
the conventional estimate of the Lorentz factor.
We discuss the possibility that the maximum energy of detected photons 
does not provide the best indicator of the minimum Lorentz factor.
The pair-break energy is also easily missed 
because the steepening of the spectral index is relatively small
(less than $(\beta-1)/4 \sim 0.3$).
Then the current observations still allow smaller Lorentz factor than the
previous estimates.
We have applied the cutoff smearing effect to 
GRB080916C and suggested that the observations
are consistent with the Lorentz factor
of $\sim 600$ (and even smaller value) that is slightly below the 
previous result of $\sim 900$.

The Lorentz factor of the outflow is likely to have a large dispersion in the internal shock model.
When $A$ is large, the resulting spectrum has lower pair-break energy where the smearing starts, and vice versa. 
This suggests that dispersion of observed sub-pulses may have some clues to estimate the Lorentz factor of 
the outflow.   

There are differences between the results of the numerical calculation and those of an analytical one, although the 
analytical calculation is a good approximation for small fluctuation of an initial Lorentz factor. 
The steepening is small in the low energies for large $A$, compared with the analytical calculation. 
However, numerical values approach analytical values in the high energies by the effect of multiple collisions.

Although we have demonstrated that the smeared cutoff behavior is characterized by steepened power-law spectra, 
there are a lot of uncertainties in the predicted high-energy spectra even in the internal shock model 
(see Sections \ref{subsec:other} and \ref{subsec:radiative}). 
(1) We have fixed $\varepsilon_{b}$, $\epsilon_{e}$ and $\beta$, but changing either of them for each collision 
easily changes the resulting spectra. Even the power-law approximation for accelerated electrons may not be good, because 
it depends on the unknown acceleration mechanism. (2) There may be an additional high-energy component coming from e.g., the 
inverse-Compton scattering and/or baryon-initiated electromagnetic cascade processes. (3) The cutoff may originate from 
other processes such as the maximum energy of accelerated electrons. There is also some uncertainty in the high energies 
where attenuation by the CIB is important due to the ambiguity of the CIB. 
Although the smeared cutoff spectrum may easily change if we consider the above effects, it is commonly expected that 
the obvious cutoff cannot be seen for time-integrated spectra of bursts with 
many sub-pulses in light curves, as long as 
we believe the internal shock model. Such a smearing effect is also expected for other high-energy features. For example,
a part of the authors provided the recipe to diagnose the mechanism of prompt emission for individual sub-pulses and 
discussed the pair annihilation bump \citep{Murase:2007ya}. We can expect the smearing effect if photospheric emission is 
also caused by the internal shock dissipation. Even high-energy neutrino spectra (see for high-energy neutrinos, e.g., \cite{m07}, and references there in) can be modified since the magnetic 
field will be weaker at larger emission radii that can affect some of the previous results \citep{2006ApJ...640L...9A}. 
The recently launched \textit{Fermi} can be useful for testing the smeared cutoff spectrum. However, it seems that 
we still need many $>$ GeV photons for further investigations. Therefore, not only \textit{Fermi} but also current 
Cherenkov detectors such as MAGIC or planned low-energy threshold Cherenkov detectors such as CTA and 5@5 will be important. 
Once we see high-energy events by these larger area telescopes, good photon statistics will allow us to study the 
high-energy behavior of prompt emission well. 

Recently, Zhang \& Pe'er argued that the emission radius of GRB080916C should
be larger than $r \gtrsim {10}^{15}$~cm. Our estimate based on the 
internal shock model suggests $r \gtrsim {10}^{14.5}$~cm. 
Since the true Lorentz factor in the multi-zone scenario can be smaller than in the one-zone scenario, even smaller emission radii are possible. They also argued that the pure fireball model 
may not avoid the photospheric emission as the relic thermal emission. 
Note that, even if the outflow is initially Poynting-dominated, the internal shock model may still be
one of the viable models as long as a significant fraction of the magnetic energy can be converted into kinetic energy \citep{2003ApJ...596.1104V}. 
Recently, models that GRB emission is powered by dissipation of the
Poynting flux energy within the outflow have been discussed more and
more \citep{2003astro.ph.12347L, 2009MNRAS.394L.117N}. 
Future observations of broadband spectra that are well time resolved will give us very crucial clues to the realistic GRB prompt emission model.  

\acknowledgments
The authors thank the anonymous referee for useful suggestions that
improved this paper. J.A. thanks Y. Sendouda for fruitful discussions. 
This research was supported by a Grant-in-Aid for the Global COE Program 
The Next Generation of Physics, Spun from Universality and Emergence 
from Ministry of Education, Culture, Sports, Science and Technology (MEXT). 
J.A. and K.M. are supported by Grants-in-aAid for Japan Society for the Promotion of Science (JSPS) 
Fellows from MEXT. 
K.T. is supported in part by Monbukagaku-sho Grant-in-Aid
for the global COE programs,
Quest for Fundamental Principles in the Universe:
from Particles to the Solar System and the Cosmos,
at Nagoya University.
This research was supported by Grant-in-Aid for Scientific Research on Priority Areas from MEXT 
19047004 (K.I. and S.N.), Grant-in-Aid for Scientific Research (S) 19104006 (S.N.) from JSPS, 
Grant-in-Aid for young Scientist(A) 21684014 (K.I.), and Grant-in-Aid for young 
Scientists(B) 18740147 (K.I.) and 19740139 (S.N.) from JSPS.
The numerical calculations were carried out on Compaq Alpha Server
ES40 at Yukawa Institute for Theoretical Physics, Kyoto University.

\appendix

\section{Two-Shell Interaction}
\label{subsec:two}
We consider the situation that a rapid shell collides with a slower shell. We assume two shells merge after collision. 
This assumption is valid when the merged shell becomes cool immediately (i.e., the cooling time by emission is shorter than 
the dynamical time in which a shock wave crosses a shell). We calculate the physical quantities of the merged shell defined in the source frame, Lorentz factor $\Gamma_{m}$, 
the internal energy $E_{\rm{int}}$, number density of protons $n_{p,m}$, thickness of the shell $l_{m}$, and area $\Sigma_{m} \simeq 4 \pi r^2$ using initial values of Lorentz factor $\Gamma$, 
density of protons $n_p$, thickness $l$, and area $\Sigma \simeq 4 \pi r^2$ (i.e., volume of the shell is $V\simeq \Sigma l$) of two shells. Physical quantities of a rapid (slower) shell is denoted by the 
subscript "$r$" ("$s$"). The kinetic energy of two shells converts to the internal energy when two shells collide and shock waves occur. Using conservation 
of momentum and energy, we can calculate the velocity and the internal energy of the merged shell: 
\begin{equation}
\Gamma_{m}\simeq \sqrt{\frac{m_{r}\Gamma_{r}+m_{s}\Gamma_{s}}{m_{r}/\Gamma_{r}+m_{s}/\Gamma_{s}}},
\end{equation}
\begin{equation}
E_{\rm{int}}=m_{r}c^2(\Gamma_{r}-\Gamma_{m})+m_{s}c^2(\Gamma_{s}-\Gamma_{m}),
\end{equation}
where $m=m_{p}n_p V$ ($m_{p}$ is the proton mass). This internal energy is radiated immediately. We calculate the energy spectrum of the emission in Section 
\ref{subsec:radiation}. 

Two shock waves occur when shells merge, a forward shock and a reverse shock. The velocity of a forward 
(reverse) shock $\Gamma_{\rm{fs}}$ ($\Gamma_{\rm{rs}}$) is written as $\Gamma_{\rm{fs}} \simeq \Gamma_m \sqrt{(1+2\Gamma_{m}/\Gamma_{s})/(2+\Gamma_{m}/\Gamma_{s})}$ 
$ \left( \Gamma_{\rm{rs}} \simeq \Gamma_m \sqrt{(1+2\Gamma_{m}/\Gamma_{r})/(2+\Gamma_{m}/\Gamma_{r})} \right) $ \citep{1995ApJ...455L.143S,1997ApJ...490...92K}. These shocks 
compress the initial shells and the thickness of the merged shell $l_{m}$ is given by
\begin{equation}
l_{m} = l_{s} \frac{\beta_{\rm{fs}} - \beta_{m}}{\beta_{\rm{fs}} - \beta_{s}} + l_{r} \frac{\beta_{m} - \beta_{\rm{rs}}}{\beta_{r}-\beta_{\rm{rs}}}.
\end{equation}
For simplicity, we assume the density of the merged shell becomes homogeneous, although there is the contact 
discontinuity in fact. We use the averaged density written as:
\begin{equation}
n_{p,m} = \frac{n_{p,r} l_r + n_{p,s} l_{s}}{l_{m}}.
\end{equation}

\section{Emission from Multiple Shells}
\label{sec:multi}
We consider the multiple shells colliding with each other and the emission from the merged shell.
We can calculate the energy spectrum of the emission by applying the discussion of the previous 
section to each collision. 

We calculate the dynamics of shells in one dimension numerically \citep{1997ApJ...490...92K}. We consider 
$N$ shells 
which are labeled by an index $i$ $(i=1,\dots,N)$, where the inner shell is labeled by the larger number. 
These shells are characterized by four variables, Lorentz factor $\Gamma_i$, density $n_i$, thickness 
of a shell $l_i$, and the initial position of (inner part of ) the shell $r_i$. We determine the initial 
position so as to the length 
between the two shells $d_i(=r_i-r_{i+1}-l_{i+1})$ becomes equal (i.e., $d_i$ is same for all shells). We also assume $d_i$ is comparable to $l_i$ and set $d_i = 5 \times l_i$. 
Lorentz factors are highly relativistic value and distributed randomly following the log-normal distribution. 
The log-normal distribution is defined by $\Gamma_0$ and the amplitude of the fluctuation $A$ as follows: 
\begin{eqnarray}
\label{eq:lognormal}
\ln \frac{\Gamma-1}{\Gamma_0-1} = A x,~~ P(x) d x= 
\frac{e^{-x^2/2}}{\sqrt{2\pi}} d x .
\end{eqnarray}
Its mean value is $\exp(\ln(\Gamma_0-1)+A^2/2)$ and its variance is 
$(\exp(A^2)-1)\exp(2\ln(\Gamma_0-1)+A^2) (\equiv \sigma_\Gamma^2)$.
At $A < 1$, we have $(\Gamma-\Gamma_0)/\Gamma_0 \simeq A x$, then $\Gamma_0$ becomes the mean 
value and $\sigma_{\Gamma}/\Gamma_0 \simeq A$. At $A > 1$, we are in the high amplitude regime. The specific values of parameters are described in the caption of the figures. Note that we chose the width of the 
shell in order to set the variance time at about $10$ or $100$ ms. 
At $t=0$ the shells are at the initial positions.
These shells expand spherically with highly relativistic speed and density decreases in proportion to $r^{-2}$, where 
$r$ is the radius of the expanding shell. 
Strictly, radial velocity spread causes a gradual spread of radial width of shells at large radius. 
Although the shell spreading would be important at large radii, 
we do not consider the shell spreading 
in this paper for the ease of understanding. 
A rapid shell catches up with a slower one and both collide with each other.  We 
assume two shells merge after collision and the shell becomes cold immediately via emission. We can calculate the 
internal energy which occurs in the $j$th collision $E_{\rm{int},j}$ and the energy spectrum of photons $dn_j/d\varepsilon$ (see Section \ref{subsec:radiation}). We also 
calculate the position $R_j$ and the time $t_j$.  An observer at $D$ away from the central source will begin to detect 
the emission at a time $t_{j,\rm ob} = [D-R_j]/c+t_j$. 

Shells expand and collide with other shell one after another. This process continues until there is no rapid shell behind the slower shell or 
all shells merge into one shell. Note that we do not consider the interstellar medium (ISM). Most shells collide at the small radius where the effect 
of the ISM is negligible when we consider the large value of $A$. The shell collides at the large radius and the effect of the ISM is important 
and the shell would be decelerated when we consider low $A$.

We have the time-integrated energy spectrum 
$dn_{\rm{sum}}/d\varepsilon = \Sigma^{N_{\rm coll}}_{j=1} 
dn_j/d\varepsilon$, where $N_{\rm coll}$ is 
the number of collisions. We focus on 
the time-integrated spectrum in this study. 
We define the duration $T$ of the GRB by the time we detect 95 \% of the total energy. We define the time-averaged luminosity $L_{\gamma}$ by 
$L_{\gamma}  = {(1+z)}^2
\left[ \int d \varepsilon 
\varepsilon \left(dn_{\rm{sum}}/d\varepsilon \right) \right] / T$.

\clearpage

\end{document}